\documentclass[showpacs,preprintnumbers,amssymb]{revtex4}
\usepackage{epsfig}

\def\beq{\begin{eqnarray}}
\def\eeq{\end{eqnarray}}
\def\nnb{\nonumber}
\newcommand{\be}{\begin{equation}}
\newcommand{\ee}{\end{equation}}
\newcommand{\bea}{\begin{eqnarray}}
\newcommand{\eea}{\end{eqnarray}}
\newcommand{\ba}{\begin{array}}
\newcommand{\ea}{\end{array}}
\begin{document}
\title{Lepton Flavor Violating $\tau^- \to \mu^- V^0$ Decays
in the Two Higgs Doublet Model III}
\author{Wenjun Li}
\email{liwj24@163.com}
 \affiliation{ Department of Physics, Henan
Normal University, XinXiang, Henan, 453007, P.R.China \nnb \\ Kavli
Institute for Theoretical Physics China, CAS, Beijing 100190, China}
\author{ Yanqin Ma, Gongwei Liu, Wei Guo}
 \affiliation{ Department of Physics,
Henan Normal University, XinXiang, Henan, 453007, P.R.China}

\begin{abstract}
In this paper, the lepton flavor violating $\tau^- \to \mu^-
V^0(V^0=\rho^0,\phi,\omega)$ decays are studied in the framework of
the two Higgs doublet model(2HDM) III. We present a computation of
the the $\gamma-$, $Z$ penguin and box diagrams contributions, and
make an analysis of their impacts. Our results show that, among the
$\gamma-$ penguins, the penguins with neutral Higgs in the loop are
very larger than those with charged Higgs in the loop. We find that
the model parameter $\lambda_{\tau\mu}$ is tightly constrained at
the order of $O(10^{-3})$ and the branching ratios of these decays
are available at the experiment measure. With the high luminosity,
the B factories have considerable capability to find these LFV
processes. On the other hand, these processes can also provide some
valuable information to future research and furthermore present the
reliable evidence to test the 2HDM III model.

\end{abstract}


\maketitle

\noindent
\section{\bf Introduction}
    The flavor physics is always the hot subject in particle
physics. Recently, with rapid development of neutrinos
experiment\cite{superK}, the lepton flavor violation(LFV) processes
of charged-lepton sector have attracted many people's attention. In
the standard model(SM), the LFV processes are forbidden. Hence, the
LFV decays are expected to be a powerful probe to many extensions of
the SM with new LFV source and/or new particles.

    The LFV $\tau$ decays have become a seeking goal in
experiment. Due to the comparability of $e^+e^-\to b\bar{b}$ and
$e^+e^-\to \tau^- \tau^-$ cross section ($\sigma \sim 0.99nb$)
around the $\Upsilon(4s)$ energy region, large events of $\tau$
leptons are available at BaBar and Belle (${\cal
L}_{BaBar}~470fb^{-1}, {\cal L}_{Belle} ~710fb^{-1}$). And now the
tau pairs production has attained the reach of $10^{-9}$. The tau
factory has performed the experimental search for the tau radiative
decays and $\tau \to 3 l$ decays, as well as $\tau \to l V^0$
decays\cite{bebar}. The current experimental upper limits of the
$\tau^- \to \mu^- \rho^0(\phi,\omega)$ decays with 543$fb^{-1}$ of
data at Belle laboratory are~\cite{belle}: \bea &&{\cal B}(\tau^-
\to \mu^- \rho^0)<6.8\times10^{-8},\,\,\,\,90\%CL
\nnb \\
&&{\cal B}(\tau^- \to \mu^- \phi)<1.3 \times 10^{-7},\,\,\,\,90\%CL
\nnb
\\&&{\cal B}(\tau^- \to \mu^- \omega)<8.9 \times
10^{-8}, \,\,\,\,90\%CL
 \eea

There are also lots of theoretical researches on $\tau \to l V^0$
decays in many possible extensions of the SM. For example, Saha
$\textit{et al.}$
 have deliberated constraints on the parameters from $\tau \to l
\rho^0(\phi,K^{*0},\bar{K}^{*0})$ decays in RPV SUSY
model\cite{Saha}. Ilakovac $\textit{et al.}$ found only the ratios
of $\tau^- \to e^- \rho^0( \phi,\pi^0)$ decays reach the order of
$10^{-6}$ in models with heavy Dirac or Majorana
neutrinos\cite{Ilakovac}. The case of $\tau \to l P(V^0)$ decays in
topcolor model have been considered by Yue Chongxing $\textit{et
al}.$\cite{Yue}. Such investigations also have been presented in
MSSM and minimal susysemmetry $SO(10)$ models\cite{Fukuyama}, a
general unconstrained MSSM model\cite{Rossi} and two constrained
MSSM seesaw models\cite{Arganda} as well.

In our previous work\cite{li}, we have studied the $\tau \to \mu
P(P=\pi^0,\eta,\eta')$ decays in 2HDM model III. In this model,
there exist flavor-changing neutral currents(FCNCs) at tree level.
In order to satisfy the current experiment constrains, the
tree-level FCNCs are suppressed in low-energy experiments for the
first two generation fermions. While processes concerning with the
third generation fermions would be larger. These FCNCs with neutral
Higgs bosons mediated may produce sizable effects to the $\tau -
\mu$ transition. The $\tau \to \mu P$ decays could yield one
pseudoscalar meson from the vacuum state through the scalar and
pseudoscalar currents. Hence, this type decay could occur at the
tree level through the neutral Higgs bosons exchange. In this paper,
we extend our discussion to the case of one vector meson in the
hadronic final state. Different from pseudoscalar meson, the vector
meson is only generated through vector currents and therefore
receive no contributions of the neutral Higgs at tree level. So we
consider the effects with Higgs bosons in the loop. There are the
$\gamma-,$ $Z$ penguin and the box diagrams for the $\tau \to
\rho^0(\phi,\omega)$ decays.
 For the instance of vector meson $K^{*0}(\bar{K}^{*0})$,
  the LFV processes could occur at loop
level likewise but the additional loop at the hadronic vertex would
generate one suppressed factor. So these two decays are not
discussed in this paper. Our results suggest that, in the $\gamma-$
penguins, the contributions of penguin with neutral Higgs bosons in
the loop is greater than those of penguin with charged Higgs bosons
in the loop. The model parameter $\lambda_{\tau\mu}$ is restrained
at $O(10^{-3})$ and the decay branching ratios could as large as the
current upper limits of $O(10^{-7})$. For $\tau^- \to \mu^- PP$
processes, we will make further study in our later work.

The paper is organized as follows: In section II, we make a brief
introduction of the theoretical framework for the two-Higgs-doublet
model III. In section III, we present the decay amplitudes and the
numerical predictions for the branching ratios. Our conclusions are
listed in the last section.
\section{\bf The Two-Higgs -Doublet Model III}
As the simplest extension of the SM, the Two-Higgs-Doublet Model has
an additional Higgs doublet. In order to ensure the forbidden FCNCs
at tree level, it requires either the same doublet couple to the
\textit{u}-type and \textit{d}-type quarks(2HDM I) or one scalar
doublet couple to the \textit{u}-type quarks and the other to
\textit{d}-type quarks(2HDM II). While in the 2HDM
III\cite{cheng,2hdm3}, two Higgs doublets could couple to the
\textit{u}-type and \textit{d}-type quarks simultaneously.
Particularly, without an {\it ad hoc} discrete symmetry exerted,
this model permits flavor changing neutral currents occur at the
tree level.

The Yukawa Lagrangian is generally expressed as the following form:

\beq {\cal L}_{Y}= \eta^{U}_{ij} \bar Q_{i,L} \tilde H_1 U_{j,R} +
\eta^D_{ij} \bar Q_{i,L} H_1 D_{j,R} + \xi^{U}_{ij} \bar
Q_{i,L}\tilde H_2 U_{j,R} +\xi^D_{ij}\bar Q_{i,L} H_2 D_{j,R} \,+\,
h.c., \label{lyukmod3}
 \eeq
 where $H_i(i=1,2)$ are the two Higgs doublets.
 $Q_{i,L}$ is the left-handed
 fermion doublet, $U_{j,R}$ and
$D_{j,R}$ are the right-handed singlets, respectively. These
$Q_{i,L}, U_{j,R}$ and $D_{j,R}$ are weak eigenstates, which can be
rotated into mass eigenstates. While $\eta^{U,D}$ and $\xi^{U,D}$
are the non-diagonal matrices of the Yukawa couplings.

We can conveniently choose a suitable basis to denote $H_1$ and
$H_2$ as: \bea
 \label{base}
 H_1=\frac{1}{\sqrt{2}}\left[\left(\ba{c} 0 \\
v+\phi^0_1 \ea\right)+ \left(\ba{c} \sqrt{2}\, G^+\\
i G^0\ea\right)\right], \,\,\,\,\,\,\,\,
 H_2=\frac{1}{\sqrt{2}}
 \left(\ba{c}\sqrt{2}\,H^+\\ \phi^0_2+i A^0\ea\right),
 \eea
where $G^{0,\pm}$ are the Goldstone bosons, $H^{\pm}$ and $A^0$ are
the physical charged-Higgs boson and CP-odd neutral Higgs boson,
respectively. Its virtue is the first doublet $H_1$ corresponds to
the scalar doublet of the SM while the new Higgs fields arise from
the second doublet $H_2$.

The CP-even neutral Higgs boson mass eigenstates $H^0$ and $h^0$ are
linear combinations of $\phi_1^0$ and $\phi^0_2$ in Eq.(\ref{base}),
 \bea \label{masseigen}
H^0 & = & \phi_1^0 \cos\alpha + \phi^0_2\sin\alpha ,\,\,\, h^0  =
-\phi^0_1\sin\alpha + \phi^0_2 \cos\alpha   ,
 \eeq
 where $\alpha$ is the mixing angle.

After diagonalizing the mass matrix of the fermion fields, the
Yukawa Lagrangian becomes\cite{David}
 \bea  L_Y&=&-\overline{U}M_UU-\overline{D}M_DD
 +\frac{i}{\upsilon}\chi^0\left(
 \overline{U}M_U\gamma_5U-\overline{D}M_D\gamma_5D\right)\nnb \\
 &+&\frac{\sqrt{2}}{\upsilon}
\chi^-\overline{D}V^\dagger_{CKM}\left[M_UR-M_DL\right]U
-\frac{\sqrt{2}}{\upsilon}\chi^+
\overline{U}V_{CKM}\left[M_DR-M_UL\right]D\nnb \\
&+&\frac{iA^0}{\sqrt{2}}\left\{\overline{U}\left[\widehat{\xi}^UR-\widehat{\xi}^{U\dag}L\right]U
+\overline{D}\left[\widehat{\xi}^{D\dag}L-\widehat{\xi}^{D}R\right]D\right\}\nnb\\
 &-&\frac{H^0}{\sqrt{2}}\overline{U}\left\{\frac{\sqrt{2}}{\upsilon} M_U
 \cos\alpha+\left[\widehat{\xi}^UR+\widehat{\xi}^{U\dag}L\right]\sin\alpha\right\}U
 -\frac{H^0}{\sqrt{2}}\overline{D}\left\{\frac{\sqrt{2}}{\upsilon} M_D
 \cos\alpha+\left[\widehat{\xi}^DR+\widehat{\xi}^{D\dag}L\right]\sin\alpha\right\}D\nnb\\
 &-&\frac{h^0}{\sqrt{2}}\overline{U}\left\{-\frac{\sqrt{2}}{\upsilon} M_U
 \sin\alpha+\left[\widehat{\xi}^UR+\widehat{\xi}^{U\dag}L\right]\cos\alpha\right\}U
-\frac{h^0}{\sqrt{2}}\overline{D}\left\{\frac{\sqrt{2}}{\upsilon}
M_D
 \sin\alpha+\left[\widehat{\xi}^DR+\widehat{\xi}^{D\dag}L\right]\cos\alpha\right\}D\nnb\\
 &-&H^+\overline{U}\left[V_{CKM}\widehat{\xi}^DR-\widehat{\xi}^{U\dag}V_{CKM}L\right]D
 -H^-\overline{D}\left[\widehat{\xi}^{D\dag}V^\dagger_{CKM}L-V^\dagger_{CKM}\widehat{\xi}^UR\right]U
 \label{lyukmass}
  \eea
where U and D now are the fermion mass eigenstates and
 \bea
\hat\eta^{U,D}&=&(V_L^{U,D})^{-1}\cdot \eta^{U,D} \cdot
V_R^{U,D}=\frac{\sqrt{2}}{v}M^{U,D}(M^{U,D}_{ij}=\delta_{ij}m_j^{U,D}),
\label{diag}\\
\hat\xi^{U,D}&=&(V_L^{U,D})^{-1}\cdot \xi^{U,D} \cdot V_R^{U,D}
\label{neutral},
 \eea
 where $V_{L,R}^{U,D}$ are
the rotation matrices acting on up and down-type quarks, with left
and right chiralities respectively. Thus
$V_{CKM}=(V_L^U)^{\dag}V_L^D$ is the usual Cabibbo-Kobayashi-Maskawa
(CKM) matrix. In general, the matrices $\hat\eta^{U,D}$ of
Eq.(\ref{diag}) are diagonal, while the matrices $\hat\xi^{U,D}$ are
 non-diagonal which could induce scalar-mediated FCNC. Seen
from Eq.({\ref{lyukmass}}), the coupling of neutral Higgs bosons to
the fermions could generate FCNC parts.
  For the arbitrariness of definition for $\xi^{U,D}_{ij}$ couplings,
we can adopt the rotated couplings expressed $\xi^{U,D}$ in stead of
$\hat{\xi}^{U,D}$ hereafter.

In this work, we use the Cheng-Sher ansatz\cite{cheng} \be
\xi^{U,D}_{ij}=\lambda_{ij} \,\frac{\sqrt{m_i m_j}}{v} \label{sher}
\ee
 which ensures that the FCNCs within the first two
generations are naturally suppressed by small fermions masses. This
ansatz suggests that LFV couplings involving the electron are
 suppressed, while LFV transitions involving muon and tau
are much less suppressed and may lead to some loop effects which are
promising to be tested by the future B factory
 experiments. In Eq.(\ref{sher}), the parameter $\lambda_{ij}$ is complex
 and $i,j$ are the generation indexes. In this study, we shall discuss the phenomenological
applications of the type III 2HDM.

\section{\bf The discussion for $\tau^- \to \mu^- V^0$ decays }
As we have mentioned above, one vector meson could not be generated
from the vacuum state through the scalar and/or pseudoscalar
currents. In 2HDM model III, the neutral Higgs bosons mediated tree
and penguin diagrams have no contributions to $\tau^- \to \mu^- V^0$
processes. Accordingly, their decay amplitudes acquire contributions
from the $\gamma-, Z-$ penguin and box diagrams.
 Comparing to the $\tau \to \mu
P$decays, in addition to neutral Higgs bosons, the penguin with
charged Higgs bosons in the loop also contribute to these decays. We
will make a detail analysis of their effects in the later
paragraphs. The penguin diagrams at the quark level pertinent to
these decays are list in Fig.1.

\begin{figure} \label{feyn}
\begin{minipage}[t]{7in}\vskip-6cm
\epsfig{file=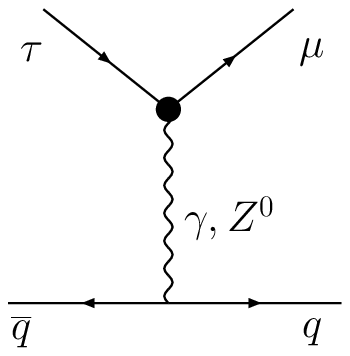,width=5in}\vskip-8cm \hspace{8cm}
\caption{The $\gamma$ and $Z^0$ penguin diagrams for $\tau^- \to
\mu^- q \bar{q}$ decay, where the neutral and charged Higgs bosons
are in the loop. }
\end{minipage}
\end{figure}

  The amplitudes could be factorized into leptonic vertex corrections
  and hadronic parts described with hadronic matrix elements.
  In dealing with hadronic matrix elements, we take the generalized
factorization approach and write the hadronic matrix elements as
$<V|\bar{q}\gamma_\mu q|0>= -m_Vf_V\varepsilon_\mu^*$ with the decay
constant $f_V$. The quark contents of $\rho^0$ meson are chosen as
$\rho^0 = \frac{1}{\sqrt{2}}(-u\bar{u}+d\bar{d})$. For the vector
$\phi-\omega$ meson system, we employ the ideal mixing scheme
between $\phi(1020)$ and $\omega(782)$ which is supported by
existing data: $\phi = -s \bar{s},\omega =
\frac{1}{\sqrt{2}}(u\bar{u}+d\bar{d})$\cite{xiaoz}. Then, the total
amplitudes could be expressed as: \bea {\cal M}(\tau^- \to \mu^- V^0
)&=&{\cal M}_\gamma+{\cal M}_Z+{\cal
M}_{box} \nnb \\
%refer to 07-7-17formular:taulv.tex
{\cal M}_\gamma(\tau^-\to \mu^- V^0) &=& \frac{i
\alpha^2_wS_w^2}{2m^2_w}\cdot \bar{\mu} \cdot [F1_\gamma \cdot L
+F2_\gamma \cdot R+F3_\gamma\cdot\gamma_\rho L
+F4_\gamma\cdot\gamma_\rho R] \cdot \tau \nnb
\\&&\otimes\langle V^0|\frac{2}{3}\bar{u}\gamma^\rho u
-\frac{1}{3}\bar{d}\gamma^\rho d -\frac{1}{3}\bar{s}\gamma^\rho
s|0\rangle\nnb \\
 {\cal M}_Z&=&\frac{i\alpha_w^2}{8m^4_W}
  \cdot \bar{\mu} \cdot \biggl[
  F_1^z \cdot L+F_2^z \cdot R+
  F_3^z \cdot\gamma_\rho L +F_4^z
   \cdot\gamma_\rho R]\cdot \tau \nnb \\&& \otimes
  \langle V^0|g^u_V(\bar{u}\gamma^\rho
 u)-g^d_V(\bar{d}\gamma^\rho
d)-g^s_V(\bar{s}\gamma^\rho s ) |0\rangle \nnb \\
{\cal M}_{box}&=&\frac{i\alpha_w^2}{m^4_W}  \cdot \bar{\mu}\cdot
\biggl[
  F_1^{box}\cdot L+F_2^{box} \cdot R+
  F_3^{box} \cdot\gamma_\rho L
  +F_4^{box} \cdot \gamma_\rho R+
    F_5^{box}\cdot i\sigma_{\rho\lambda}L
    +F_6^{box}\cdot i\sigma_{\rho\lambda}R\biggl]
  \cdot \tau \nnb \\&& \otimes \langle V^0|\bar{u}\gamma^\rho
  u-\bar{d}\gamma^\rho d-\bar{s}\gamma^\rho  s |\rangle \label{amp}
  \eea
where ${\cal M}_\gamma,{\cal M}_Z$ and ${\cal M}_{box}$ are the
amplitudes of the $\gamma-$ penguin, Z penguin and box diagrams. The
relevant auxiliary functions are listed in Appendix.

    In our calculation, the input
parameters are the Higgs masses, mixing angle $\alpha$,
 $|\lambda_{ij}|$ and their phase angles $\theta_{ij}$.
 Given the constraints from the current experiment permits and
 theoretical considerations\cite{li,dyb,Atwood,csh,Rodolfo,zhuang,Rozo},
 we assume \vskip-1cm
 \bea m_{H^\pm}&=&200GeV,\,\,\, m_{H^0}=160GeV, \,\,\,m_{h^0}=115GeV,\,\,\,
  m_{A^0}=120GeV,\,\,\,\alpha=\pi/4,\,\,\nnb \\
    |\lambda_{uu}|&=&150,\,\,\,
|\lambda_{dd}|=120,\,\,\,|\lambda_{\tau\tau}|=10,\,\,\,\,|\lambda_{tt}|=|\lambda_{tc}|=|\lambda_{ut}|=0.03,\,\, \nnb \\
|\lambda_{ss}|&=&|\lambda_{bb}|=|\lambda_{db}|=|\lambda_{bs}|=100,\,\,
\theta = \pi/4,\,\, \eea where the Higgs masses satisfy the relation
$115GeV\leq m_{h^0}< m_{A^0}< m_{H^0}\leq
200GeV$\cite{dyb,csh,Rodolfo}, and the absolute value of
$\lambda_{tt}\cdot \lambda_{bb}$ is approximate to
three\cite{csh,zhuang}.

Using the above parameters, we could get the contributions of three
diagrams to these decays. As we expected, the contributions of box
diagrams are $O(10^{-25})$ order or so which are very smaller than
those of $\gamma-$ and $Z-$ penguins. Hence, we neglect the box
diagrams contributions. We have studied the relation of branching
ratio and $\lambda_{\tau\mu}$. The computation indicate that the
variation of $\theta_{\tau\mu}$, the phase angle of parameters
$\lambda_{\tau\mu}$, does almost not affect the values of branching
ratios. So we take $\theta_{\tau\mu}= \pi/4$ as literatures do.

The Fig.2 gives the total penguin contributions denoted by the solid
line. We denote the $\gamma-$ penguin and the $Z-$ penguin
contributions by the dash line and the dot line, respectively. Due
to the suppressed factor $O(1/m^2_Z)$ from the Z propagator, the $Z$
penguin contributions are supposed to be lower than those of the
$\gamma-$ penguin. These decay amplitudes have common leptonic
parts, so the differences of decay amplitudes mainly come from the
hadronic parts. For the similar contents of $\rho^0$ and $\omega$,
the curves of $\tau^- \to \mu^- \rho^0$ and $\tau^- \to \mu^-
\omega$ decays display similar trend, namely, their $Z$ penguin
contributions are lower one order than those of their $\gamma-$
penguin. However, for $\tau^- \to \mu^- \phi$ decay, the magnitudes
of $Z$ penguin are close to the $\gamma-$ penguin contributions.

The relations of ratios versus $|\lambda_{\tau\mu}|$ are also
presented in Fig.2, where the horizon lines denote the experimental
upper limits. Evidently, one can see from Fig.2 that these branching
ratios rise with the increase of $|\lambda_{\tau\mu}|$. We have got
the constraints on $|\lambda_{\tau\mu}|$ from the experimental data,
which are list in Table.I. It is obviously that the parameter
$|\lambda_{\tau\mu}|$ is restrained at the order of $O(10^{-3})$.
The $|\lambda_{\tau\mu}|$ constraints for $\tau^- \to \mu^-
\rho^0(\omega)$ are little severe than that of $\tau^- \to \mu^-
\phi$ decay. The bounds of $|\lambda_{\tau\mu}|$ from different
phenomenological considerations
\cite{li,sher,Martin,Rodolfo,Zhou,Cotti} are demonstrated in Tab.I,
too. Comparing the values of $\lambda_{\tau\mu}$ in Tab.I, one can
see that our constraint is stringenter than the limits in
literatures.

  \begin{table}[tb]\label{ql}
\caption{Constraints on the $\lambda_{\tau\mu}$ from $\tau^- \to
\mu^- \rho^0(\phi,\omega)$ decays in the 2HDM III. }
\begin{center}
\begin{tabular}
{ccc}
\hline \hline Decay modes\,\,\,& Bounds on $\lambda_{\tau\mu}$ & Previous Bounds \\
\cline{1-3}  \hline
  $\tau \to \mu \rho^0$\,\,\,&$\leq 1.26\times 10^{-3}$ \,\,\, &$\lambda_{\tau\mu}\sim O(1)$ ~\cite{sher}
\\
 \cline{1-3}
 $\tau \to \mu \phi$\,\,\,&$\leq 2.45\times 10^{-3}$ \,\,\,&  $\lambda_{\tau\mu}\sim O(10)$ ~\cite{li,Martin,Zhou}, $\lambda_{\tau\mu}\sim O(10)-O(10^2)$~\cite{Rodolfo}\\
\cline{1-3}
  $\tau \to \mu \omega$\,\,\,&$\leq 1.48\times 10^{-3}$\,\,\,&$\lambda_{\tau\mu}\sim O(10^2)-O(10^3)$ ~\cite{Cotti}\\
\hline\hline
\end{tabular}
\end{center}
\end{table}

\begin{figure}[htbp] \label{fall}
\includegraphics[width=18cm]{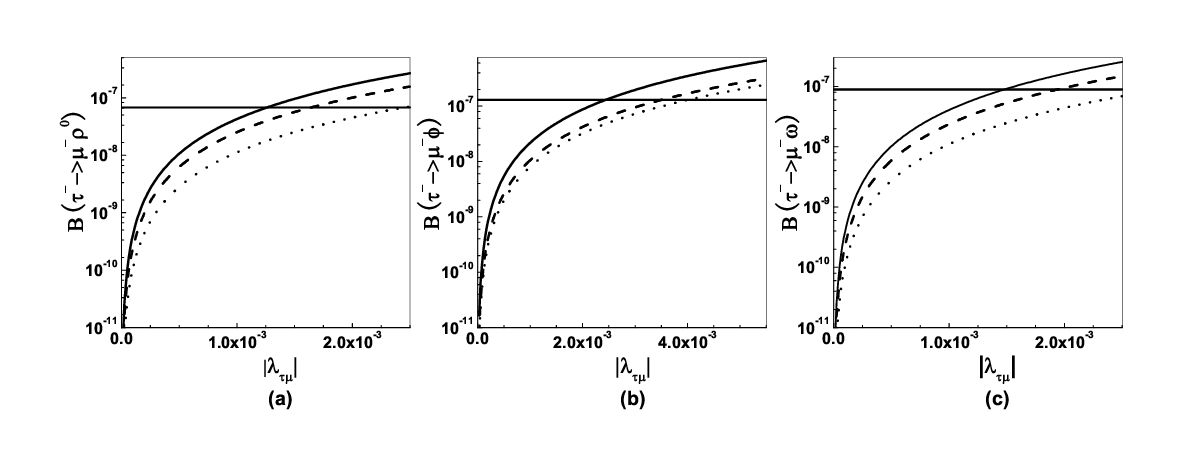}
\vskip-1cm \caption{The branching ratios versus model parameter
$|\lambda_{\tau\mu}|$ with $\theta_{\tau\mu}=\pi/4$, (a) for $\tau^-
\to \mu^- \rho^0$ decay, (b) for $\tau^- \to \mu^- \phi$ decay, and
(c) for $\tau^- \to \mu^- \omega$ decay. The solid line denotes the
total contributions; the dash line and the dot line denote the
$\gamma-$ and Z penguin contributions, respectively. The horizontal
lines are the experimental upper limits.}
\end{figure}

\begin{figure} \label{fr}
\includegraphics[width=18cm]{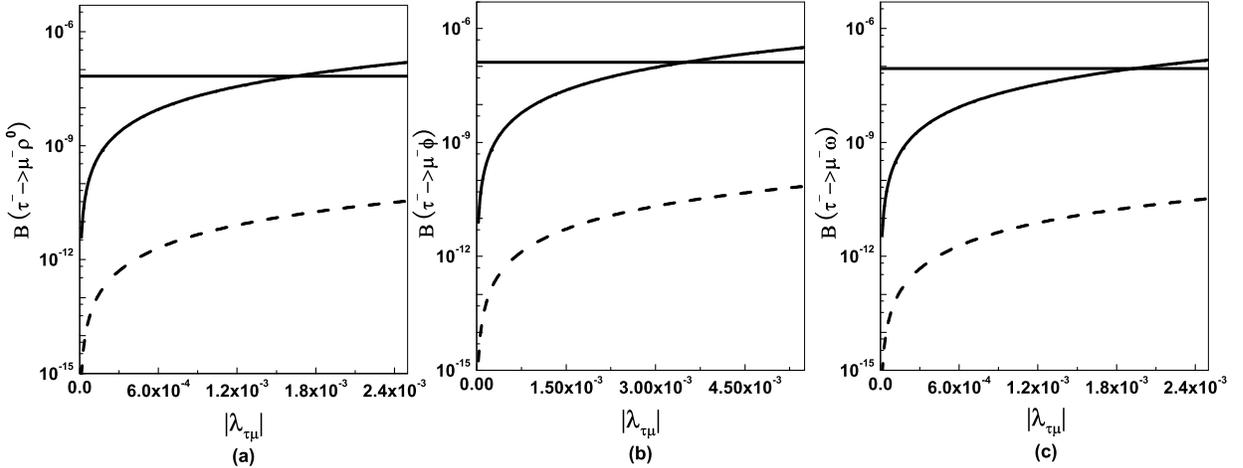}\vskip-1.5cm
\caption{The branching ratios versus model parameter
$|\lambda_{\tau\mu}|$ with $\theta_{\tau\mu}=\pi/4$, (a) for $\tau^-
\to \mu^- \rho^0$ decay (b) for $\tau^- \to \mu^- \phi$ decay, and
(c) for $\tau^- \to \mu^- \omega$ decay. The solid line denotes the
$\gamma$ penguin contributions; the dash line and the dot line
denote the contributions of $\gamma-$ penguin with neutral Higgs
bosons in the loop and those of $\gamma-$ penguin with charged Higgs
bosons in the loop, respectively. }
\end{figure}

Now we illustrate the contributions of $\gamma-$ penguin with
neutral and charged Higgs bosons in the loop. In Fig.3, the solid
line denotes the $\gamma-$ penguin contributions, the dash line and
the dot line denote the contributions of $\gamma-$ penguin with
neutral Higgs bosons in the loop and those of $\gamma-$ penguin with
charged Higgs bosons in the loop, respectively. Apparently, the
contributions of $\gamma-$ penguin with neutral Higgs in the loop
are quite higher by nearly four order magnitudes than those of
$\gamma-$ penguin with charged Higgs in the loop. And the dot line
and the solid line coincide with each other for three decays. As a
result, the contributions of $\gamma-$ penguin with neutral Higgs in
the loop are dominated one.

\begin{figure} \label{fZ}
\includegraphics[width=18cm]{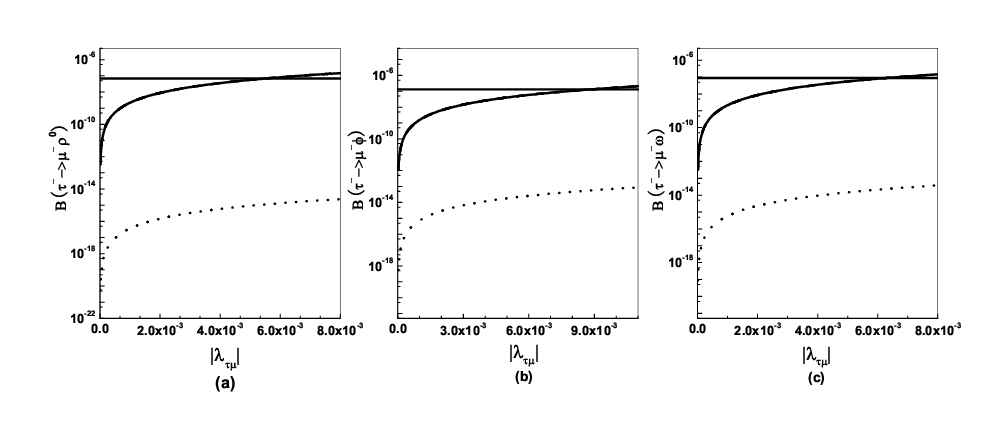}\vskip-1.5cm
\caption{The branching ratios versus model parameter
$|\lambda_{\tau\mu}|$ with $\theta_{\tau\mu}=\pi/4$, (a) for $\tau^-
\to \mu^- \rho^0$ decay (b) for $\tau^- \to \mu^- \phi$ decay, and
(c) for $\tau^- \to \mu^- \omega$ decay. The solid line denotes the
Z penguin contributions; the dash line and the dot line denote the
contributions of $Z$ penguin with charged Higgs bosons in the loop
and those of $Z$ penguin with neutral Higgs bosons in the loop,
respectively}
\end{figure}

The contributions of $Z$ penguin with charged and neutral Higgs
bosons in the loop are demonstrated in figure 4. The solid line
denotes the $Z$ penguins contributions, the dash line and the dot
line denote the contributions of $Z$ penguin with charged Higgs
bosons in the loop and those of $Z$ penguin with neutral Higgs
bosons in the loop, respectively. Unlike the case of $\gamma$
penguin, the contributions of $Z$ penguin with neutral Higgs in the
loop are rather smaller by nearly eight order magnitudes than those
of $Z$ penguin with charged Higgs in the loop. So the contributions
of $Z$ penguin with neutral Higgs in the loop are subordinate one.
In a word, the $\gamma-$ penguin with neutral Higgs in the loop
plays a main role in these decays.

\section{\bf Conclusion}
In summary, we have calculated the branching ratios of $\tau^-
 \to \mu^- \rho^0(\phi,\omega)$ decays in the model III 2HDM.
Comparing to the $\tau^-\to \mu^- P$ decays, besides the neutral
higgs bosons in the loop, an additional charged Higgs boson in the
loop offer contributions to $\tau^-\to \mu^- V^0$ decays. The
impacts of the $\gamma-$ penguin, $Z-$ penguin and those of two
types Higgs in loop are formulated. It is concluded that the
$\gamma-$ penguin with neutral Higgs bosons in loop are dominated in
the $\gamma-$ penguin, while the $Z-$ penguin with charged Higgs
bosons in loop mainly contributes to the $Z-$ penguins. Our work
suggests that the parameter $|\lambda_{\tau\mu}|$ is constrained at
the order of $O(10^{-3})$. And in the rational parameters space, the
$Br(\tau^-\to \mu^- V^0)$ can reach the experimental values. With
the experiment luminosity increasing, these LFV decays are available
to the collider's measure capability. Our study is hoped to supply
good
 information for the future experiment and explore the structure of
the 2HDM III model.

\section*{\large\bf Appendix}
For simplicity, we only list the amplitude of $\gamma-$ penguin.

 The amplitude of
  $\gamma-$ penguin diagrams is
  \bea
  {\cal M}_\gamma(\tau^-\to \mu^- V^0) &=&
\frac{i \alpha^2_wS_w^2}{2m^2_w}\cdot  \bar{\mu}\cdot
[F1_\gamma\cdot L +F2_\gamma\cdot R  +F3_\gamma\cdot  \gamma_\rho L
  +F4_\gamma\cdot \gamma_\rho R  ]\cdot \tau \nnb
\\&&\otimes\langle V^0|\frac{2}{3}\bar{u}\gamma^\rho u
-\frac{1}{3}\bar{d}\gamma^\rho d -\frac{1}{3}\bar{s}\gamma^\rho
s|0\rangle. \eea Where the auxiliary functions $F_\gamma$ are
written as: \bea F1_\gamma&=& \frac{m_\tau \sqrt{m_\tau m_\mu}\cdot
 }{k^2} \cdot \int_0^1 dx \int_0^{1-x}dy \left\{\biggl(-
 \frac{m_\tau \lambda^*_{\tau\mu} \lambda_{\tau\tau}\cdot x}{S_{c}(x,y,m^2_{H^-},x_{tc})}
+\frac{m_\tau \lambda^*_{\tau\mu} }{2} \sum_{i} J_i\times x \biggl
)\cdot p_1^\rho \right.\nnb
\\
&& \left. + \biggl(-
 \frac{m_\tau \cdot
  \lambda^*_{\tau\mu}\lambda_{\tau\tau}\cdot x}{S_{c}(x,y,m^2_{H^-},x_{tc})}+\frac{1}{2}\sum_{i}K_i\times y\biggl)\cdot
p_2^\rho\right\}, \\
F2_\gamma&=&\frac{m_\tau \sqrt{m_\tau m_\mu}}{k^2} \cdot \int_0^1 dx
\int_0^{1-x}dy \left\{\biggl(- \frac{ m_\mu\cdot
\lambda^*_{\tau\mu}\lambda_{\tau\tau}\cdot
y}{S_{c}(x,y,m^2_{H^-},x_{tc})}+\frac{m_\tau \cdot
\lambda_{\tau\mu}}{2} \cdot \sum_{i} J^*_i\times x \biggl)\cdot
p_1^\rho \right. \nnb
\\
&&\left.+ \biggl(- \frac{m_\mu\cdot
\lambda^*_{\tau\mu}\lambda_{\tau\tau}\cdot
y}{S_{c}(x,y,m^2_{H^-},x_{tc})}+\frac{1}{2}\cdot \sum_{i}K^*_i\times
y\biggl)\cdot p_2^\rho\right\},
 \eea \vskip-0.5cm
 \bea
F3_\gamma&=& \frac{m_\tau \sqrt{m_\tau m_\mu}}{k^2} \cdot \left\{
\frac{1}{(m^2_\tau- m^2_\mu)}\cdot \int_0^1 dx \biggl[m_\mu
\cdot\lambda^*_{\tau\mu} \lambda_{\tau\tau}(x-1)\ln \frac{
S_{a}(x,x_{tc})}{S_{b}(x)}+\frac{1}{2}M_i\biggl]+N+\int_0^1 dx
\int_0^{1-x}dy Q_i\right\},
 \nnb \\ \\
F4_\gamma&=&\frac{m_\tau \sqrt{m_\tau m_\mu}}{k^2} \cdot\left\{
\frac{1}{(m^2_\tau- m^2_\mu)}\cdot\int_0^1 dx \biggl[
\lambda^*_{\tau\mu} \lambda_{\tau\tau}(x-1)\cdot \biggl(m_\tau^2\ln
S_{a}(x,x_{tc})-m_\mu^2 \ln
S_{b}(x)\biggl)+\frac{1}{2}M^*_i\biggl]+N^*\right.\nnb
\\&&\left.+ \int_0^1 dx \int_0^{1-x}dy[
\ln \frac{S_{c}(x,y,m^2_{H^-},x_{tc})}{\mu^2}+Q^*_i ] \right\}.
 \eea
The followings are expressions of $J_i,K_i,M_i,N$ and $Q_i$. \bea
J_{H^0}&=&
\frac{\omega_s}{S^{H^0}_{c}(x,y,m^2_{H^0},x^{H^0}_{tn})},\,\,\,\,\,
J_{h^0}=\frac{\upsilon_s}{S^{h^0}_{c}(x,y,m^2_{h^0},x^{h^0}_{tn})},\,\,\,\,\,J_{A^0}=
\frac{2i
Im\lambda_{\tau\tau}}{S^{A^0}_{c}(x,y,m^2_{A^0},x^{A^0}_{tn})},\nnb
\\
K_{H^0}&=&(m_\tau \lambda^*_{\tau\mu} +m_\mu
\lambda_{\tau\mu})\times J^*_{H^0},\,\,\,\,\,K_{h^0}=(m_\tau
\lambda^*_{\tau\mu} +m_\mu \lambda_{\tau\mu})\times J^*_{h^0},
\,\,\, \,\,\,K_{A^0}= \frac{\lambda^*_{\tau\tau}( m_\mu
\lambda_{\tau\mu}-m_\tau
\lambda^*_{\tau\mu})}{S^{A^0}_{c}(x,y,m^2_{A^0},x^{A^0}_{tn})}\nnb\\
  M_{H^0}&=& \biggl[[
x(m_\tau^2 \omega^*\lambda_{\tau\mu} +m_\tau m_\mu
\omega\lambda^*_{\tau\mu}) -\omega_s
(m_\tau^2\lambda_{\tau\mu}+m_\tau m_\mu \lambda^*_{\tau\mu}) ]\ln
S^{H^0}_{a}(x,x^{H^0}_{tn})\nnb
\\&&- [
 x(m_\mu^2\omega^*\lambda_{\tau\mu}
+m_\tau m_\mu \omega\lambda^*_{\tau\mu})
-(m_\mu^2\omega^*+m_\tau^2\omega)\lambda_{\tau\mu} -m_\tau m_\mu
\omega_s \lambda^*_{\tau\mu}]\ln S^{H^0}_{b}(x)\biggl ]\nnb \\
Q_{H^0}&=& \int_0^1 dx \int_0^{1-x}dy\biggl(\omega^*
\lambda_{\tau\mu}[\ln \frac{S_c(x,y,m^2_{H^-},x_{tc})}{\mu^2}
 +\frac{m^2_\tau(x^2-x-1) -m^2_\mu y}{S_{c}^{H^0}(x,y,m^2_{H^0},x^{H^0}_{tn})}]
\nnb \\
&&+ \frac{m_\tau m_\mu \lambda^*_{\tau\mu} [(x+y)\omega-\omega_s]
 -m^2_\tau \omega \lambda_{\tau\mu}}{S_{c}^{H^0}(x,y,m^2_{H^0},x^{H^0}_{tn})}
 \biggl)\nnb \\
M_{h^0}&=&
  [
 x(m_\tau^2 \upsilon^*\lambda_{\tau\mu}
+m_\tau m_\mu  \upsilon\lambda^*_{\tau\mu}) -\upsilon_s
(m_\tau^2\lambda_{\tau\mu}+m_\tau m_\mu \lambda^*_{\tau\mu})]\ln S^{h^0}_{a}(x,x^{h^0}_{tn})\nnb \\
 &&- [
 x(m_\mu^2\upsilon^*\lambda_{\tau\mu}
+m_\tau m_\mu  \upsilon\lambda^*_{\tau\mu})
-(m_\mu^2\upsilon^*+m_\tau^2\upsilon)\lambda_{\tau\mu}-m_\tau m_\mu
\upsilon_s\lambda^*_{\tau\mu}]\ln S^{h^0}_{b}(x)\biggl]
\nnb \\
Q_{h^0}&=&\biggl([\ln \frac{S_c(x,y,m^2_{H^-},x_{tc})}{\mu^2}
 +\frac{m^2_\tau(x^2-x-1) -m^2_\mu y}{S_{c}^{h^0}(x,y,m^2_{h^0},x^{h^0}_{tn})}] \upsilon^* \lambda_{\tau\mu}
\nnb \\
&&+ \frac{m_\tau m_\mu \lambda^*_{\tau\mu}[(x+y)\upsilon
-\upsilon_s]
 -m^2_\tau \upsilon \lambda_{\tau\mu}}{S_{c}^{h^0}(x,y,m^2_{h^0},x^{h^0}_{tn})}
 \biggl)\nnb \\
 M_{A^0}&=& \biggl[[x(m_\tau^2\lambda_{\tau\tau}^*\lambda_{\tau\mu}
+m_\tau m_\mu \lambda_{\tau\tau} \lambda^*_{\tau\mu})-
2iIm\lambda_{\tau\tau}(m_\tau m_\mu
\lambda^*_{\tau\mu}-m_\tau^2\lambda_{\tau\mu}) ]\ln
S^{A^0}_{a}(x,x^{A^0}_{tn})\nnb \\&&-
 [x (m_\mu^2 \lambda_{\tau\tau}^*\lambda_{\tau\mu}
+m_\tau m_\mu  \lambda_{\tau\tau}\lambda^*_{\tau\mu})
 - (m_\mu^2\lambda_{\tau\tau}^*-m_\tau^2 \lambda_{\tau\tau})
 \lambda_{\tau\mu}
 -2i m_\tau m_\mu \lambda^*_{\tau\mu}Im\lambda_{\tau\tau}
]\ln S^{A^0}_{b}(x)\biggl]\nnb \\
 Q_{A^0}&=&\biggl([
\ln \frac{S^{A^0}_c(x,y,m^2_{A^0},x^{A^0}_{tn})}{\mu^2}
+\frac{m^2_\tau(x^2-x-1)-m^2_\mu y
}{S^{A^0}_c(x,y,m^2_{A^0},x^{A^0}_{tn})}] \lambda^*_{\tau\tau}
\lambda_{\tau\mu}
\nnb \\
&&+\frac{m^2_\tau \lambda_{\tau\tau} \lambda_{\tau\mu} +m_\tau
m_\mu\lambda^*_{\tau\mu}[(x+y)\lambda_{\tau\tau}
-2iIm\lambda_{\tau\tau} ] }{S^{A^0}_{c}(x,y,m^2_{A^0},x^{A^0}_{tn})}
\biggl) \nnb \\
N&=&\frac{1}{2} \lambda_{\tau\mu}(\omega^* +\upsilon^*
+\lambda^*_{\tau\tau})\nnb \\
\omega&=&(\lambda_{\tau\tau}\sin^2 \alpha +\sin \alpha \cos
\alpha),\,\,\,\,
\upsilon=(\lambda_{\tau\tau}\cos^2 \alpha -\sin \alpha \cos \alpha)\nnb\\
\omega_s&=&2\sin^2\alpha Re\lambda_{\tau\tau}+\cos 2 \alpha,\,\,\,\,
\upsilon_s=2\cos^2\alpha Re\lambda_{\tau\tau}-\cos 2 \alpha \nnb
\eea The integrate function expressions are : \bea
 S_{a}(x,x_{tc})&=&(x-1)(x_{tc}x-1),\,\,
\,\,\,
 S_{b}(x)=1-x ,\,\,\,
S_{c}(x,y,m^2_{H^-},x_{tc})=m^2_{H^-}[x+(x^2-x+y)x_{tc}],\,\,\, x_{tc}=\frac{m^2_\tau}{m^2_{H^-}},\nnb \\
 S^i_{a}(x,x^i_{tn})&=&(x-1)(x^i_{tn}x-1) \,\,\,\,\,
 S^i_{b}(x)= S_{b}(x),\,\,\,
S^i_{c}(x,y,m^2_i,x^i_{tn})=m^2_i[y+x+x^i_{tn}x(x-1)],\,\,x^i_{tn}=\frac{m^2_\tau}{m^2_i}
 \eea

\begin{acknowledgments}
I thank Prof.Chaoshang Huang for discussion. The work is supported
by National Science Foundation under contract No.10547110, He¡¯nan
Educational Committee Foundation under contract No.2007140007, the
Project of Knowledge Innovation Program (PKIP) of Chinese Academy of
Sciences under Grant No.KJCX2.YW.W10.

\end{acknowledgments}

\end{document}